\documentclass[12pt]{article}
\usepackage{t1enc}
\usepackage[latin2]{inputenc}
\usepackage{amsmath}
\usepackage{amssymb}
\usepackage{amsthm}

\newtheorem{thm}{Theorem}
\newtheorem{defin}[thm]{Definition}
\newtheorem{lemma}[thm]{Lemma}

\newtheorem{cor}[thm]{Corollary}
\newenvironment{prf}{\noindent\textit{Proof}.\ }{\eop}

\newcommand\Bbox{
{\unskip\nobreak\hfil\penalty 50
\hskip 1em\hbox{}\nobreak\hfil{\lower .5pt \hbox{$\Box$}}
\parfillskip=0pt \finalhyphendemerits=0 \par}
}
\newcommand\eop{\ifmmode {\hbox{\Bbox}} \else \Bbox \fi}

\newcommand{\E}{\mathrm{E}}

\newcommand{\opt}{\mathrm{opt}}
\newcommand{\cost}{\mathrm{cost}}

\title{Randomized algorithm for the k-server problem on decomposable spaces\footnote{Research is partially supported by OTKA T049398.}}

\author{Judit Nagy-Gy\"orgy\thanks{Department of Mathematics, University of
Szeged, Aradi v\'ertan\'uk tere 1, H-6720 Szeged, Hungary, email:
Nagy-Gyorgy@math.u-szeged.hu}}

\begin{document}
\frenchspacing
\maketitle
\begin{abstract}
We study the randomized $k$-server problem on metric spaces consisting of widely separated subspaces. 
We give a method which extends existing algorithms to larger spaces with the growth rate of the competitive quotients 
being at most $O(\log k)$.
This method yields $o(k)$-competitive algorithms solving the randomized $k$-server problem, for some special underlying metric spaces, e.g. HSTs of ``small'' height (but unbounded degree). HSTs are important tools for probabilistic approximation of metric spaces.
\end{abstract}

\noindent \textbf{Keywords:}  $k$-server, on-line, randomized, metric spaces.
\section{Introduction}
In the theory of designing efficient virtual memory-management algorithms,
the well studied paging problem plays a central role. Even the earliest
operation systems contained some heuristics to minimize the amount of
copying memory pages, which is an expensive operation.
A generalization of the paging problem, called
the $k$-server problem was introduced by Manasse, McGeoch and Sleator in \cite{MMS},
where the first important results were also achieved. The problem can be formulated
as follows. Given a metric space with $k$ mobile servers that occupy distinct
points of the space and a sequence of requests (points), each of the requests has
to be served, by moving a server from its current position to the requested
point. The goal is to minimize the total cost, that is the sum of the distances
covered by the servers; the optimal cost for a given sequence $\varrho$ is
denoted $\opt(\varrho)$.

An algorithm is online if it serves each request immediately when it arrives
(without any prior knowledge about the future requests).

\begin{defin}
An online algorithm $A$ is $c$-competitive if for any initial configuration $C_0$
and request sequence $\varrho$ it holds that
$$\cost(A(\varrho)) \le c\cdot \opt(\varrho) + I(C_0),$$
where $I$ is a non-negative constant depending only on $C_0$.
\end{defin}

The competitive ratio of a given online algorithm $A$ is the infimum of the
values $c$ with $A$ being $c$-competitive. The $k$-server conjecture (see \cite{MMS})
states that there exists an algorithm $A$ that is $k$-competitive for any
metric space. Manasse et al. proved that $k$ is a lower bound \cite{MMS},
and Koutsoupias and Papadimitriou showed $2k-1$ is an upper bound for any metric
space \cite{KP}.

In the randomized online case (sometimes this model is called the oblivious adversary model \cite{BB}) the competitive ratio can be defined in terms of
the expected value as follows:

\begin{defin}
A randomized online algorithm $R$ is $c$-competitive if for any initial configuration
$C_0$ and request sequence $\varrho$ we have
$$\E(\cost(R(\varrho))) \le c\cdot \opt(\varrho) + I(C_0),$$
where $I$ is a non-negative constant depending only on $C_0$ and $\E(\cost(R(\varrho)))$ denotes
the expected value of $\cost(R(\varrho))$.
\end{defin}

The competitive ratio of the above randomized algorithm is defined analogously.

In the randomized version there are more problems that are still open. The randomized
$k$-server conjecture states that there exists a randomized algorithm with
a competitive ratio $\Theta(\log k)$ in any metric space. The best known lower bound
is $\Omega(\log k/\log^2 \log k)$ \cite{BBM}. A natural upper bound is the bound $2k+1$
given for the deterministic case. By restricting our attention to metric spaces with
a special structure, better bounds can be achieved: Fiat et al. showed a lower bound
$H_k=\sum_{i=0}^k i^{-1}\approx \log k$ for uniform metric spaces \cite{pag}, which
turned out to be also an upper bound, see McGeoch and Sleator, \cite{MS}.

In this paper we also consider a restriction of the problem, namely we seek for
an efficient randomized online algorithm for metric spaces that are 
``$\mu$-HST spaces'' \cite{B1,B2} and defined as follows:

\begin{defin}
  A $\mu$-hierarchically well-separated tree ($\mu$-HST) is a metric space defined on the
  leaves of a weighted, rooted tree $T$ with the following properties:
  \begin{enumerate}
  		\item The edge weight from any node to each of its children is the same.
  		\item The edge weights along any path from the root to a leaf are decreasing by the factor 
			$\mu$ from one level to the next. The weight of an edge incident to a leaf is one.
  \end{enumerate}
\end{defin}

The $\mu$-HST spaces play an important role in the so-called metric space approximation
technique developed by Bartal \cite{B2}. Fakcharoenphol et al \cite{F} proved that 
every weighted graph on $n$ vertices can be $\alpha$-probabilistically approximated
by a set of $\mu$-HSTs, for an arbitrary $\mu>1$ where $\alpha=O(\mu \log n/\log \mu)$.

It has been shown in \cite{S} that for any
$2k$-HST with an underlying tree $T$ that has a small depth and maximum degree there
exists a $\mathrm{polylog}(k)$-competitive randomized algorithm for the $k$-server problem. 
By slightly modifying the approach of Csaba and Lodha \cite{CL} and Bartal and Mendel \cite{BM} 
we show that there exists such an algorithm for any $\mu$-HST that has a small depth and 
arbitrary maximum degree $t$, given $\mu \geq \min\{k,t\}$.

\section{Notation}

Suppose the points of a metric space can be partitioned into $t$ blocks, $B_1,\dots,B_t$,
such that the diameter of each block is at most $\delta$ and whenever $x$ and $y$ are points
of different blocks, their distance is exactly $\Delta$. Suppose also that
$\Delta/\delta = \mu \ge k$ holds. The above metric space is $\mu$-decomposable \cite{S}.

For a given request sequence $\varrho$ we denote its $i$th member by $\varrho_i$,
and the prefix of $\varrho$ of length $i$ by $\varrho_{\le i}$.

Given a block $B_s$, a request sequence $\varrho$ having only requests from $B_s$
and a number $\ell$ of servers inside $B_s$ and an algorithm $A$, let
$\cost(A_s(\ell,\varrho))$ denote the cost computed by the algorithm $A$
for these inputs and $\opt_s(\ell,\varrho)$ (in the latter case also the initial
position of the servers can be chosen). If $\varrho$ is nonempty, $\opt_s(0,\varrho)$
is defined to be infinite.

\begin{defin}\label{dem}
  The \emph{demand} of the block $B_s$ for the request sequence $\varrho$ that contains
  only requests from $\varrho$ is
  $$D_s(\varrho) := \min\{\ell\, |\, \opt_s(\ell,\varrho) + \ell\Delta = \mathop{\min}_{j}\{\opt_s(j,\varrho) + j\Delta\}\},$$
  if $\varrho$ is nonempty, otherwise it is $0$.
\end{defin}

Visually, $D_s(\varrho)$ denotes the least number of servers to be moved
into the initially empty block $B_s$ to achieve the optimal cost for the sequence $\varrho$.
We note that the behaviour of the sequence
$D_s(\varrho_1),D_s(\varrho_{\le 2}),\ldots,D_s(\varrho_{\le i}),\ldots,$ $D_s(\varrho)$
is unclear.

\section{Algorithm X}
In the rest of the paper we suppose that there exists a randomized online algorithm $A$
and a function $f$ with $\frac{f(\ell)}{\log \ell}$ being monotone increasing (constants are allowed), and we have
\begin{equation}\label{komp}
\E[\cost(A_s(\ell,\varrho))]\le f(\ell)\cdot \opt_s(\ell,\varrho) + \frac{f(\ell)\cdot\ell\delta}{\log \ell}
\end{equation}
for any $\ell$ and $s$. Having Algorithm $A$, we can define our ``shell algorithm'' X that
uses $A$ as a subroutine inside the blocks.

\subsection{The Algorithm}
The algorithm uses $A$ as a subroutine and it works in phases. Let $\varrho^{(p)}$ denote
the sequence of the $p$th phase. In this phase the algorithm works as follows:

\begin{quote}
Initially we mark the blocks that contain no servers.

When $\varrho_i^{(p)}$, the $i$th request of this phase arrives to block $B_s$,
we compute the demand $D_s(\varrho^{(p)}_{\le i})$ and the maximal demand
$$D_s^*(\varrho_i^{(p)})=\max\{D_s(\varrho_{\le j}^{(p)}) | j\le i\}$$
for this block (note that these values do not change in the other blocks).

-- If $D_s^*(\varrho_i^{(p)})$ is less than the number of servers in $B_s$ at that moment,
   then the request is served by Algorithm $A$, with respect to the block $B_s$.

-- If $D_s^*(\varrho_i^{(p)})$ becomes equal to the number of servers in $B_s$ at that
   moment, then the request is served by Algorithm $A$, with respect to the block $B_s$
   and we mark the block $B_s$.

-- If $D_s^*(\varrho_i^{(p)})$ is greater than the number of servers in $B_s$ at that
   moment, we mark the block $B_s$ and perform the following steps until we have
   $D_s^*(\varrho_i^{(p)})$ servers in that block or we cannot execute the steps
   (this happens when all the blocks become marked):

\begin{itemize}
\item Let us choose an unmarked block $B_{s'}$ randomly uniformly, and a server from this block also
  randomly. We move this chosen server to the block $B_s$ (such a move is called
  a \emph{jump}), either to the requested point, or,
  if there is already a server occupying that point, to a randomly chosen unoccupied point of $B_s$.
  If the number of servers in $B_{s'}$ becomes $D_{s'}^*(\varrho_i^{(p)})$ via this move, we mark
  that block. In both $B_s$ and $B_{s'}$ we restart algorithm $A$ from the current configuration
  of the block.
\end{itemize}

If we cannot raise the number of servers in block $B_s$ to $D_s^*(\varrho_i^{(p)})$ by
repeating the above steps (all the blocks became marked), then Phase $p+1$ is starting
and the last request is belonging to this new phase.
\end{quote}

Our main result is the following:

\begin{thm}\label{fo}
  Algorithm X is $c\cdot\log k\cdot f(k)$-competitive for some constant $c$.
\end{thm}

In the following two subsections we will give an upper bound for the cost of Algorithm X 
and several lower bounds for the optimal cost in an arbitrary phase. 
The above theorem easily follows from these.

For convenience we modify the request sequence $\varrho$ in a way that
does not increase the optimal cost and does not decrease the cost of any
online algorithm, hence the bounds we get for this modified sequence will
hold also in the general case. The modification is defined as follows: we
extend the sequence by repeatedly
requesting the points of the halting configuration of a (fixed)
optimal solution.
We do this till $\sum_{s=1}^t D_s^*(\varrho_{\le i}^{(u)})$ becomes $k$.
Observe that the optimal cost does not change via this transformation, and
any online algorithm works the same way in the original part of the sequence
(hence online), so the cost computed by any online algorithm is at least the
original computed cost.

\subsection{Upper bound}
In the first step we prove an auxiliary result. We recall from
\cite{CP} that an online matching problem is defined similarly to the online
$k$-server problem with the following two
differences:
\begin{enumerate}
	\item Each of the servers can moved only once;
	\item The number of the requests is at most $k$, the number of the servers.
\end{enumerate}
For any phase $p$ of Algorithm X we can associate the following matching
problem MX. The underlying metric space of the matching problem is a
finite uniform
metric space that has the blocks $B_s$ as points and a distance $\Delta$ between
any two different points. Let $\hat{D}_s(p)$ denote the number of servers that
are in the block $B_s$ just at the end of phase $p$. Now in the associated
matching problem we have $\hat{D}_s(p-1)$ servers originally occupying the
point $B_s$. During phase $p$, if some value $D^*_s$ increases, we make a number
of requests in point $B_s$ for the associated matching problem: we make the
same number of requests that the value $D^*_s$ has been increased with.

We also associate an auxiliary matching algorithm (AMA) on this structure as
follows. Suppose $D^*_s$ increases at some time, causing jumps. These jumps
are corresponding to requests of the associated matching problem; AMA satisfies
these requests by the servers that are corresponding to those involved in
these jumps. 

Let $\hat{D}_s(p)$ denote the number of servers in block $B_s$ just after
phase $p$. If $p$ is not the last phase, let $\varrho^{(p)+}$ denote the
request sequence we get by adding the first request of phase $p+1$ to
$\varrho^{(p)}$. Now we have
\begin{equation}\label{plusz1}
D_s^*(\varrho^{(p)})\le \hat{D}_s(p)\le D_s^*(\varrho^{(p)+})
\end{equation}
and in all block but at most one we have equalities there (this is the block
that causes termination of the $p$th phase).

Denote
\begin{equation}\label{mp}
m_p := \sum_{s=1}^{t} \max\{0,\hat{D}_s(p)-\hat{D}_s(p-1)\}.
\end{equation}
Since the auxiliary metric space is uniform, the optimal cost is
$\Delta m_p$.
\begin{lemma}[Csaba, Pluh\'ar, \cite{CP}]
\label{prop-logkcdotDeltam_p} The expected cost of AMA
is at most $\log k\cdot \Delta m_p$.
\end{lemma}

\begin{lemma}\label{felso}
The expected cost of Algorithm X in the $p$th phase is at most
\begin{equation*}
f(k)\left(\sum_{s=1}^t \opt_s(D_s(\varrho^{(p)+}),\varrho^{(p)}) +  \Delta\left(\sum_{s=1}^t D_s(\varrho^{(p)+})-k\right)\right) +
\end{equation*}
\begin{equation*}
+ \Delta m_p(f(k)\log k + f(k) + \log k) + \Delta\frac{f(k)}{\log k}.
\end{equation*}
\end{lemma}
\begin{prf}
Consider the $p$th phase of a run of Algorithm X on the request sequence
$\varrho$ and let $\tau$ denote the associated run of AMA.

Let $B_s$ be a block in which some request arrives during this phase.
For the sake of convenience we omit the index $s$ of the block: let
$\opt(\ell,\varrho'):=\opt_s(\ell,\varrho')$, $\hat{D}_{p-1}:=\hat{D}_s(p-1)$
and $\hat{D}_{p}:=\hat{D}_s(p)$.

Denote $\varrho^{(p)}$ the restiction of $\varrho$ to $B_s$. While the block
is unmarked, only jump-outs can happen from this block; let these jump-outs happen just before the $r_1$th, $\dots,r_{d^-}$th request of $\varrho^{(p)}$,
respectively. After the block has been marked, only jump-ins can happen;
let these happen when the $r_{d^-+1}$th,$\dots,r_{d^-+d^+}$th request arrives,
respectively (for any given request there can be more preceeding jumps).
Denote $\sigma_i = \varrho_{r_i}\ldots\varrho_{r_{i+1}-1}$ (where $\varrho_{r_0}$ is the first and 
$\varrho_{r_{d^-+d^++1}-1}$ is the last request of the phase in $B_s$), 
and let $k_{s,i}:=k_i$ be the number of servers in $B_s$ during $\sigma_i$.
Observe that the demand at the $r_i$th request is exactly $k_i$. Finally,
let $\ell_i$ denote the demand occuring at the $r_i-1$th request if this request falls into 
the $p$th phase; otherwise let $\ell_i=0$.

A jump-in to the block satisfies the last request, hence there is no
server movement inside the block during a jump. The expected cost of non-jump
movements in this block (this is called the \emph{inner cost}) is, applying
(\ref{komp}), at most
\begin{eqnarray}\label{felt}
\sum_{i=0}^{d^-+d^+}\E[A_{s}(k_i,\sigma_i)|\,\tau] &\le& \sum_{i=0}^{d^-+d^+} \big(f(k_i)\opt(k_i,\sigma_i) + \frac{k_i \cdot f(k_i)}{\log k_i} \delta\big) \nonumber\\
&\le& f(k)\sum_{i=0}^{d^-+d^+}\opt(k_i,\sigma_i) + \delta\sum_{i=0}^{d^-+d^+}\frac{k_i \cdot f(k_i)}{\log k_i}.
\end{eqnarray}
We bound the right side of (\ref{felt}) piecewise.
Summing up till the jump-out just before the last:
\begin{equation}\label{ki}
\sum_{i=0}^{d^--1}\opt(k_{i},\sigma_i)\le \sum_{i=0}^{d^--1}\opt(k_{d^-},\sigma_i)\le \opt(k_{d^-},\varrho_{\le r_{d^-}}^{(p)}).
\end{equation}
From the last jump-out till the last jump-in:
\begin{eqnarray}
\sum_{i=d^-}^{d-1}\opt(k_{i},\sigma_i) &\le& \sum_{i=d^-}^{d-1}\opt(\ell_{i+1},\sigma_i) \nonumber\\
&=& \sum_{i=d^-}^{d-1}\big( \opt(\ell_{i+1},\sigma_i) + \opt(\ell_{i+1},\varrho_{\le r_i}^{(p)}) - \opt(\ell_{i+1},\varrho_{\le r_i}^{(p)})\big) \nonumber\\
&\le& \sum_{i=d^-}^{d-1}\big(\opt(\ell_{i+1},\varrho_{< r_{i+1}}^{(p)}) - \opt(\ell_{i+1},\varrho_{\le r_i}^{(p)})\big) \label{korr}\\
&\le& \sum_{i=d^-}^{d-1}\big(\opt(k_{i+1},\varrho_{< r_{i+1}}^{(p)}) + (k_{i+1}-\ell_{i+1})\Delta - \nonumber\\
&& \qquad\qquad -\ \opt(k_{i},\varrho_{\le r_i}^{(p)}) - (k_{i}-\ell_{i+1})\Delta\big)\nonumber\\
&\le& \sum_{i=d^-}^{d-1}\big(\opt(k_{i+1},\varrho_{\le r_{i+1}}^{(p)}) - \opt(k_{i},\varrho_{\le r_i}^{(p)}) + (k_{i+1}-k_{i})\Delta\big)\nonumber\\
&=& \opt(k_{d},\varrho_{\le r_{d}}^{(p)}) - \opt(k_{d^-},\varrho_{\le r_{d^-}}^{(p)}) +  (k_{d}-k_{d^-})\Delta,\label{be}
\end{eqnarray}
where $d=d^-+d^+$. Inequality (\ref{korr}) comes from Definition \ref{dem},
since the demand of $\varrho_{<r_{i+1}}$ is $\ell_{i+1}$ and the demand of
$\varrho_{\leq r_i}$ is $k_i$.

Since $k_d\geq D(\varrho^{(p)})$, analogously we get
\begin{eqnarray}
\opt(k_{d},\sigma_d) &\le& \opt(D(\varrho^{(p)}),\sigma_d)\le\nonumber\\
&\le& \opt(D(\varrho^{(p)}),\varrho^{(p)}) - \opt(D(\varrho^{(p)}),\varrho_{\le r_d}^{(p)})\le\nonumber\\
&\le& \opt(D(\varrho^{(p)+}),\varrho^{(p)}) + (D(\varrho^{(p)+})-D(\varrho^{(p)}))\Delta - \nonumber\\
&& -\ \opt(k_{d},\varrho_{\le r_d}^{(p)}) - (k_{d}-D(\varrho^{(p)}))\Delta \nonumber\\
&=& \opt(D(\varrho^{(p)+}),\varrho^{(p)}) - \opt(k_{d},\varrho_{\le r_d}^{(p)}) + (D(\varrho^{(p)+}) - k_d)\Delta.\label{vege}
\end{eqnarray}
Observe that if the request causing termination of phase $p$, that is the
first request of phase $p+1$, then $D(\varrho^{(p)})=D(\varrho^{(p)+})$ holds.

Summing up the right hand sides of (\ref{ki}), (\ref{be}) and (\ref{vege}) we
get
\begin{equation}
\sum_{i=0}^{d^-+d^+}\opt(k_i,\sigma_i) \le \opt(D(\varrho^{(p)+}),\varrho^{(p)}) + (D(\varrho^{(p)+}) - k_{d^-})\Delta,
\end{equation}
and substituting this to the right hand side of (\ref{felt}) we get that the
expected inner cost in $B_s$ is at most
\begin{equation}\label{blokk}
 f(k)\left(\opt(D(\varrho^{(p)+}),\varrho^{(p)}) + (D(\varrho^{(p)+}) - k_{d^-})\Delta\right) + \sum_{i=0}^{d^-+d^+}\frac{k_i \cdot f(k_i)}{\log k_i}\delta.
\end{equation}
On the other hand,
\begin{equation}\label{szerver}
D(\varrho^{(p)+})-k_{d^-} = (D(\varrho^{(p)+}) - \hat{D}_p) + (\hat{D}_p-k_{d^-}),
\end{equation}
where we know that $D(\varrho^{(p)+}) = D_s(\varrho^{(p)+})$, and
$(\hat{D}_p-k_{d^-})$ is the number of jump-ins into this block.

We can bound the sum of the expressions of the form $\frac{k_i \cdot f(k_i)}{\log k_i}\delta$ as follows:
\begin{equation}\label{plusz}
\sum_{s=1}^{t}\sum_{i=0}^{d_s^-+d_s^+}\frac{f(k)}{\log k} k_{s,i}\delta \le 
\sum_{s=1}^{t}\sum_{i=0}^{d_s^+}\frac{f(k)}{\log k} k\delta \le (|\tau|+1)\frac{f(k)}{\log k} k\delta,
\end{equation}
where $|\tau|$ is the number of the jumps during the phase. This comes from
the fact that we have only $k$ servers,
hence the sum of $k_s$ (the number of servers in the target block of
the jump) and $k_{s'}$ (where the given server jumps from) is still at most
$k$. We also remark that
\begin{equation}\label{elsoben}
|\tau| \le k.
\end{equation}
Now we bound the cost of the jumps. Let $T$ be the set of the potential jumps,
and the length of the jump sequence $\eta$. Applying Proposition~\ref{prop-logkcdotDeltam_p} we get that

\begin{equation}
\E[\eta\Delta] = \sum_{\tau \in T}P(\tau)|\tau|\Delta \le \log k \cdot \Delta  m_p \label{ugras}
\end{equation}

Summing up the results (\ref{blokk}), (\ref{szerver}), (\ref{plusz}) and
(\ref{ugras}) for all the blocks we get the following bound for the expected
cost of Algorithm X:
\begin{eqnarray}
&& \sum_{s=1}^t \sum_{i=0}^{d_s^-+d_s^+}\E[A_{s}(k_i,\sigma_i) + \eta\Delta] \label{fkorlat}\\
&=& \sum_{\tau \in T}\left(P(\tau)\sum_{s=1}^t \sum_{i=0}^{d_s^-+d_s^+}\E[A_{s}(k_i,\sigma_i)|\,\tau]\right) +  \Delta\E[\eta] \label{e1}\\
&\le& \sum_{\tau \in T}P(\tau)f(k)\sum_{s=1}^t \opt(D_s(\varrho^{(p)+}),\varrho^{(p)+}) + \label{e2}\\
&& + \sum_{\tau \in T}P(\tau)f(k)\sum_{s=1}^t(D(\varrho^{(p)+}) - \hat{D}_p)\Delta + \label{e3}\\
&& + \sum_{\tau \in T}P(\tau)f(k)|\tau|\Delta + \sum_{\tau \in T}P(\tau)(|\tau|+1)\frac{f(k)}{\log k} k\delta + \Delta\E[\eta]\\
&\le& f(k)\left(\sum_{s=1}^t \opt_s(D_s(\varrho^{(p)+}),\varrho^{(p)}) + \Delta\sum_{s=1}^t D_s(\varrho^{(p)+})-k\Delta + \log k \cdot \Delta  m_p\right)\nonumber\\[6pt]
&& + \Delta\left(\frac{f(k)}{\log k} m_p\log k + \frac{f(k)}{\log k} + m_p\log k\right) \nonumber,
\end{eqnarray}
if we apply $\sum_{\tau \in T}P(\tau)=1$ in (\ref{e1}) and (\ref{e2}), and $k\delta\le \Delta$ in (\ref{e3}).
\end{prf}

\subsection{Analyzing the optimal cost}
Consider an optimal solution of the $k$-server problem. Let $C_s^*(\varrho)$ be the maximal number of servers in $B_s$ of this optimal solution during phase $p$ and $C_s(\varrho)$ be the number of servers in $B_s$ of the optimal solution at the end $p$. If we duplicate the first request of every phase but the very
first one, and consider $\varrho^{(p)+}$ instead of $\varrho^{(p)}$, the
optimal cost does not change.

Denote 
$$ma_p^* := \sum_{s=1}^t (C_s^*(\varrho^{(p)+})-C_s(\varrho^{(p-1)+})).$$

From the definitions above it is clear that the optimal cost is at least
\begin{equation}\label{adv}
\opt(\varrho^{(p)+}) \ge \sum_{s=1}^t \opt_s(C_s^*(\varrho^{(p)+}),\varrho^{(p)+}) + \Delta\cdot ma_p^*.
\end{equation}

\begin{lemma}\label{also1}
$$\opt(k,\varrho^{(p)+}) \ge \sum_{s=1}^t \opt(D_s(\varrho^{(p)+}),\varrho^{(p)+}) + \Delta\left(\sum_{s=1}^t D_s(\varrho^{(p)+})-k\right).$$
\end{lemma}
\begin{prf}
From Definition \ref{dem} we have
\begin{eqnarray*}
&& \sum_{s=1}^t \left( \opt_s(C_s^*(\varrho^{(p)+}),\varrho^{(p)+}) + \Delta(C_s^*(\varrho^{(p)+})-C_s(\varrho^{(p-1)+}))\right)\ge\nonumber\\
&& \sum_{s=1}^t \left(\opt(D_s(\varrho^{(p)+}),\varrho^{(p)+}) 
+ \Delta(D_s(\varrho^{(p)+}) - C_s(\varrho^{(p-1)+}))\right)\nonumber
\end{eqnarray*}
Since $\sum_{s=1}^{t} C_s(\varrho^{(p-1)+}) = k$, the statement follows by (\ref{adv}).
\end{prf}

\begin{lemma}\label{also2}
The optimal cost is at least
$$\frac{1}{6}\Delta \sum_{p>1} m_p.$$
\end{lemma}
\begin{prf}
Let $\varrho^{(p)*}$ be the subsequence of $\varrho^{(p)}$ which we get by
omitting each request that arrives to a block $B_s$ after the demand of
that given block reaches $D_s^*(\varrho^{(p)+})$ (note that $\varrho^{(p)*}$
is not neccessarily a prefix of $\varrho^{(p)}$). Now we have two cases:
first, if
$D_s^*(\varrho^{(p)+}) > C_s(\varrho^{(p-1)+})$ holds, then by Definition \ref{dem}
\begin{eqnarray}
&&\opt_s(C_s^*(\varrho^{(p)+}),\varrho^{(p)+}) + \Delta(C_s^*(\varrho^{(p)+})-C_s(\varrho^{(p-1)+})) \nonumber\\
&\ge& \opt_s(C_s^*(\varrho^{(p)+}),\varrho^{(p)*}) + \Delta(C_s^*(\varrho^{(p)+})-C_s(\varrho^{(p-1)+})) \nonumber\\
&\ge& (\opt_s(D_s^*(\varrho^{(p)+}),\varrho^{(p)*}) + \Delta \left( 0,D_s^*(\varrho^{(p)+}) - C_s(\varrho^{(p-1)+}))\right)\nonumber\\ 
&\ge& \Delta \max\{0,\hat{D}_s(\varrho^{(p)}) - C_s(\varrho^{(p-1)+}))\}.\nonumber
\end{eqnarray}
Otherwise it holds that
$$\max\{0,\hat{D}_s(\varrho^{(p)}) - C_s(\varrho^{(p-1)+}))\}=0,$$
and also obviously
$$\opt_s(C_s^*(\varrho^{(p)+}),\varrho^{(p)+}) + \Delta(C_s^*(\varrho^{(p)+})-C_s(\varrho^{(p-1)+}))\ge 0.$$
From both cases we get
\begin{eqnarray}
&\opt_s(C_s^*(\varrho^{(p)+}),\varrho^{(p)+}) + \Delta(C_s^*(\varrho^{(p)+})-C_s(\varrho^{(p-1)+}))\quad \ge&\label{b1}\\[3pt]
&\Delta \max\{0,\hat{D}_s(\varrho^{(p)}) - C_s(\varrho^{(p-1)+}))\}.&\nonumber
\end{eqnarray}
Now since
$\mathop\sum\limits_{s=1}^t\hat{D}_s(\varrho^{(p)}) =
 \mathop\sum\limits_{s=1}^t C_s(\varrho^{(p-1)+}) = k$, it also holds that
\begin{eqnarray}
\mathop\sum\limits_{s=1}^t\Delta \max\{0,\hat{D}_s(\varrho^{(p)}) - C_s(\varrho^{(p-1)+}))\}&=\nonumber\\
\mathop\sum\limits_{s=1}^t\frac{1}{2}\Delta |\hat{D}_s(\varrho^{(p)}) - C_s(\varrho^{(p-1)+}))|.\label{sum-half}
\end{eqnarray}
Summing the cost of the jumps that the optimal solution performs we get
\begin{equation} 
\Delta\sum_{s=1}^t |C_s(\varrho^{(p)+})-C_s(\varrho^{(p-1)+})| \le 2\cdot \opt(k,\varrho^{(p)+}).\label{b2}
\end{equation}
Note that the factor of $2$ comes from the fact that each jump appears twice
on the left hand side. Now summing up (\ref{b1}), (\ref{sum-half}) and
(\ref{b2}) we have
\begin{eqnarray}
&&4\cdot \opt(k,\varrho^{(p)+}) \nonumber\\
&\ge& \Delta \sum_{s=1}^t \big( |\hat{D}_s(\varrho^{(p)}) - C_s(\varrho^{(p-1)+}))| + |C_s(\varrho^{(p)+})-C_s(\varrho^{(p-1)+})|\big)\nonumber\\
&\ge&  \Delta \sum_{s=1}^t |\hat{D}_s(\varrho^{(p)}) - C_s(\varrho^{(p)+}))|.\label{b3}
\end{eqnarray}
Now summing (\ref{b1}) relativized to phase $p$ and (\ref{b3}) relativized to
phase $p-1$ we get that
\begin{eqnarray}
&& 2\cdot \opt(k,\varrho^{(p)+}) + 4\cdot \opt(k,\varrho^{(p-1)+}) \nonumber\\
&\ge& \Delta \sum_{s=1}^t \big(|\hat{D}_s(\varrho^{(p)}) - C_s(\varrho^{(p-1)+}))| + |\hat{D}_s(\varrho^{(p-1)}) - C_s(\varrho^{(p-1)+}))|\nonumber\\
&\ge& \Delta \sum_{s=1}^t |\hat{D}_s(\varrho^{(p)}) - \hat{D}_s(\varrho^{(p-1)}))| = \Delta m_p,
\end{eqnarray}
and the statement follows.
\end{prf}

\begin{lemma}\label{also3}
If the $p$th phase is not the last one, then $\opt(k,\varrho^{(p)+}) \ge \Delta$.
\end{lemma}
\begin{prf}
Similarly to the proof of Lemma~\ref{also2},
\begin{eqnarray}
&&\opt_s(C_s^*(\varrho^{(p)+}),\varrho^{(p)+}) + \Delta(C_s^*(\varrho^{(p)+})-C_s(\varrho^{(p-1)+})) \nonumber\\
&\ge& \opt_s(C_s^*(\varrho^{(p)+}),\varrho^{(p)*}) + \Delta(C_s^*(\varrho^{(p)+})-C_s(\varrho^{(p-1)+})) \nonumber\\
&\ge& (\opt_s(D_s^*(\varrho^{(p)+}),\varrho^{(p)*}) + \Delta \left( 0,D_s^*(\varrho^{(p)+}) - C_s(\varrho^{(p-1)+}))\right)\nonumber\\ 
&\ge& \Delta \max\{0,\hat{D}_s(\varrho^{(p)}) - C_s(\varrho^{(p-1)+}))\},\nonumber
\end{eqnarray}
so we get that the optimal cost is at least
\begin{equation}
\sum_{s=1}^t\big(\opt_s(D_s^*(\varrho^{(p)+}),\varrho^{(p)*}) + \Delta(D_s^*(\varrho^{(p)+}) - C_s(\varrho^{(p-1)+}))\big).
\end{equation}
Now since $\sum_{s=1}^t C_s(\varrho^{(p-1)+}) = k$, moreover if the $p$th
phase is not the last one, then $\sum_{s=1}^t D_s^*(\varrho^{(p)+}) > k$ also
holds; applying these we get the statement.
\end{prf}

\subsection{Proof of Theorem \ref{fo}}
Now we are able to prove the theorem about competitiveness of Algorithm X.

\begin{prf}[Theorem \ref{fo}] If we apply (\ref{elsoben}) to the first phase
and write $k$ instead of $m_p\log k$, then Lemmas \ref{felso}, \ref{also1}, \ref{also2} and \ref{also3} give that
\begin{eqnarray*}
\E(\cost(KA(\varrho))) &\le& f(k)\opt(k,\varrho) + \\[3pt]
&&6 \cdot \opt(k,\varrho)\big(f(k)\log k +f(k)+ \log k\big) +\\[3pt]
&&\Delta \frac{f(k)}{\log k}\\
&\le& f'(k)\cdot \opt(k,\varrho) + \frac{f'(k)\cdot k\Delta}{\log k},
\end{eqnarray*}
where $f'(k) = f(k)(6\log k + 8)$.
\end{prf}

At this point we make a few remarks:
\begin{enumerate}
	\item If the starting configuration of the online problem coincides with
 		the starting configuration of an optimal solution, then it holds that
 		$\E(KA(\varrho)) \le c\cdot f(k)\cdot \log k \cdot \opt(k,\varrho)$.
	\item It is a bit more natural to require $\Delta\ge \delta M$ to hold,
		where $M$ is the size of the greatest block. If additionally $M < k$
 		holds, we get a better competitive ratio.
\end{enumerate}

\subsection{Corollaries}
Starting from the MARKING algorithm \cite{pag} and iterating Theorem~\ref{fo}
we get the following result:
\begin{cor}
There exists a $(c_1\log k)^h$-competitve randomized online algorithm on
any $\mu$-HST of height $h$ (here $\mu\geq k$), where $c_1$ is a constant.
Consequently, when $h<\frac{\log k}{\log c_1 + \log\log k}$, this algorithm
is $o(k)$-competitive.
\end{cor}
Considering only metric spaces having at most $t<k$ blocks, it is enough to
require $\mu > t$. Substituting this to Lemma~\ref{felso} we get a competitive
ratio of $c\cdot\log b\cdot f(k)$, what (applying the above corollary) gives
us the following result:
\begin{cor}
Suppose the metric space is a $\mu$-HST having only degrees of at most $b$.
Then there exists a $c_2^h$-competive randomized online algorithm, where
$h$ is the depth of the tree and $c_2=c_2(b)$ is some constant.
\end{cor}

\section{Further questions}
In the field of online optimization the concept of buying extra resources is also
investigated \cite{CINSW}. The quantity
$\min_\ell\{\opt_s(\ell,\varrho) + \ell\Delta\}$ can be seen as the optimal
cost of a model where one has to buy the servers, for a cost of $\Delta$ each.
This problem on uniform spaces was studied in \cite{CINSW}.
In this case $D_s(\varrho)$ is the number of servers bought in an optimal
solution. Now considering to sequence $\varrho_i$, the behaviour of the
associated sequence $D_s(\varrho_i)$ is unclear at the moment. It is an
interesting question whether the above sequence is monotonically increasing,
or does it hold that $|D_s(\varrho)_i - D_s(\varrho)_{i+1}|\le 1$ for each $i$.
The presented proofs would substantially simplify in both cases.

Another interesting question is that whether the $\log k$ factor in the
competitive ratio per level of the HST is unavoidable, or an overall
competitive ratio of $\Theta(\log k)$ holds for any HST.
\\ 
\\
\textbf{Acknowledgement.} The author wish to thank B\'ela Csaba for his guidance and suggestions, furthermore P\'eter Hajnal, Csanád Imreh and András Pluh\'ar for their valuable remarks.

\end{document}